\documentclass[journal]{IEEEtran}

\usepackage{flushend}
\usepackage{mathrsfs}
\usepackage{color}
\usepackage{multirow}
\usepackage[flushleft]{threeparttable}

\ifCLASSINFOpdf
   \usepackage[pdftex]{graphicx}
\else
  \usepackage[dvips]{graphicx}
\fi

\usepackage{amsmath}
\usepackage{makecell}

\usepackage{subcaption} %support subfigure

\begin{document}

\title{Demonstration of a Single-chip Dual-polarization Sideband-separation SIS mixer at 2 mm Band}

\author{Wenlei~Shan,~\IEEEmembership{Member,~IEEE,}
        Shohei~Ezaki,~\IEEEmembership{}
        and~Yoshinori~Uzawa~\IEEEmembership{}% <-this % stops a space
\thanks{Manuscript submitted June 1, 2026.}
\thanks{The work is partly supported by the Japan Society for the Promotion of Science(JSPS) KAKENHI under Grant Number 23K20871, 22H04955 and 22H04939.}
\thanks{Wenlei Shan, Shohei Ezaki, and Yoshinori Uzawa are with National Astronomical Observatory of Japan(NAOJ), Osawa 2-21-1, Mitaka, 181-8588, Tokyo, Japan (e-mails: wenlei.shan@nao.ac.jp;shohei.ezaki@nao.ac.jp;y.uzawa@nao.ac.jp).  Wenlei Shan and Yoshinori Uzawa are also with the Graduate University for Advanced Studies (SOKENDAI).}% <-this % stops a space
}

% The paper headers
\markboth{Journal of \LaTeX\ Class Files,~Vol.~XX, No.~X, JUNE~2026}%
{Shell \MakeLowercase{\textit{et al.}}: Bare Demo of IEEEtran.cls for IEEE Journals}

% make the title area
\maketitle

% As a general rule, do not put math, special symbols or citations
% in the abstract or keywords.

\begin{abstract}
Large-format heterodyne focal plane arrays require highly integrated receiver architectures while maintaining low noise and sufficient sideband rejection, which remains challenging for superconducting SIS mixers at millimeter wavelengths. In this work, we demonstrate, for the first time, a monolithic dual-polarization sideband-separating superconductor-insulator-superconductor (SIS) mixer operating at 2 mm wavelengths (125--163 GHz) on a silicon-on-insulator substrate. The integrated SIS mixer achieves a sideband rejection ratio exceeding 10 dB over the 4--8 GHz intermediate-frequency (IF) band across most of the radio-frequency (RF) range, with a minimum single-sideband (SSB) receiver noise temperature as low as $\sim$60 K. As a key building block of the hybrid planar integration (HPI) architecture, this result verifies the feasibility of highly integrated large-format heterodyne focal plane arrays, providing a practical path toward substantially expanding the field of view of millimeter- and submillimeter-wave radio telescopes.
\end{abstract}

% Note that keywords are not normally used for peerreview papers.

\IEEEpeerreviewmaketitle

\section{Introduction}

\IEEEPARstart{L}{arge}-format SIS heterodyne arrays are essential for future millimeter- and submillimeter-wave astronomy because they combine extremely high spectral resolution with a wide field of view. Future radio telescopes are expected to provide degree-scale focal planes capable of accommodating heterodyne focal plane arrays with thousands of pixels \cite{booth2024key,kawabe2016new}. Large-format heterodyne arrays are also scientifically important for millimeter- and submillimeter-wave interferometers such as the Atacama Large Millimeter/submillimeter Array (ALMA) \cite{carpenter2019alma}, where highly compact frontend assemblies are required to fit within cartridge-type receiver structures. However, current SIS receiver technology is typically limited to only several tens of pixels with sparse spacing \cite{groppi2011coherent}, far from the scale required for next-generation instruments.

The fundamental limitation of current SIS array technology is that large-format arrays cannot be practically realized by assembling discrete single-pixel modules. This becomes particularly severe for dual-polarization sideband-separating (2SB) receivers, where each pixel requires numerous mechanically and electrically complex interconnections for radio-frequency (RF), local-oscillator (LO), and intermediate-frequency (IF) signals, together with DC and magnetic-field biasing. As a result, simply scaling conventional waveguide-based single-pixel modules cannot provide either large-format arrays or highly compact focal-plane integration. Previous efforts to improve integration density include one-dimensional arrays based on conventional waveguide SIS mixers on quartz substrates \cite{groppi2006supercam}, as well as more aggressive miniaturization using superconducting planar circuits on silicon membranes \cite{barrueto2022ccat,westig2011balanced}, where components traditionally implemented by metal waveguides were replaced by planar circuits. However, these approaches did not provide a systematic architecture for scalable large-format arrays.

In sharp contrast to SIS heterodyne arrays, superconducting direct-detection cameras, such as microwave kinetic inductance detectors (MKIDs) and transition-edge sensor (TES) arrays, routinely achieve $10^{2}$--$10^{3}$ pixels per unit at millimeter wavelengths \cite{dutcher2024simons,choi2022ccat}, one to two orders of magnitude more than SIS mixer arrays. The key difference is architectural scalability: direct-detection arrays are fully planarized using superconducting thin-film circuits, whereas conventional SIS receivers still rely on mechanically assembled waveguide modules. Therefore, a similarly planarized architecture is fundamentally required for large-format SIS heterodyne arrays.

Based on this concept, we proposed hybrid planar integration (HPI) \cite{shan2018new}, in which the mixer circuitry is fully planarized in a monolithic microwave integrated circuit (MMIC), while only the LO distribution network remains as a semi-two-dimensional waveguide binary tree fabricated in a parallel layer. This architecture enables array compactness limited primarily by the feed-horn aperture, rather than by engineering constraints in interconnecting discrete components. As a result, both much higher pixel density and highly compact focal-plane integration become feasible.

The feasibility of the HPI concept was previously demonstrated using balanced SIS mixers operating at 2 mm wavelengths \cite{shan2019experimental,shan2020compact}. The next essential step toward practical astronomical applications is the implementation of 2SB functionality, which is highly desirable for ground-based observations because it suppresses atmospheric noise from the image sideband. In this work, we present the circuit design, fabrication, and experimental demonstration of a monolithic dual-polarization 2SB SIS mixer. In addition, channel crosstalk caused by substrate cavity modes, a unique issue in MMIC-based SIS mixers, is investigated and an effective mitigation method is presented.

\begin{figure*}[!t]
\centering
\includegraphics[width=0.80\textwidth,clip]{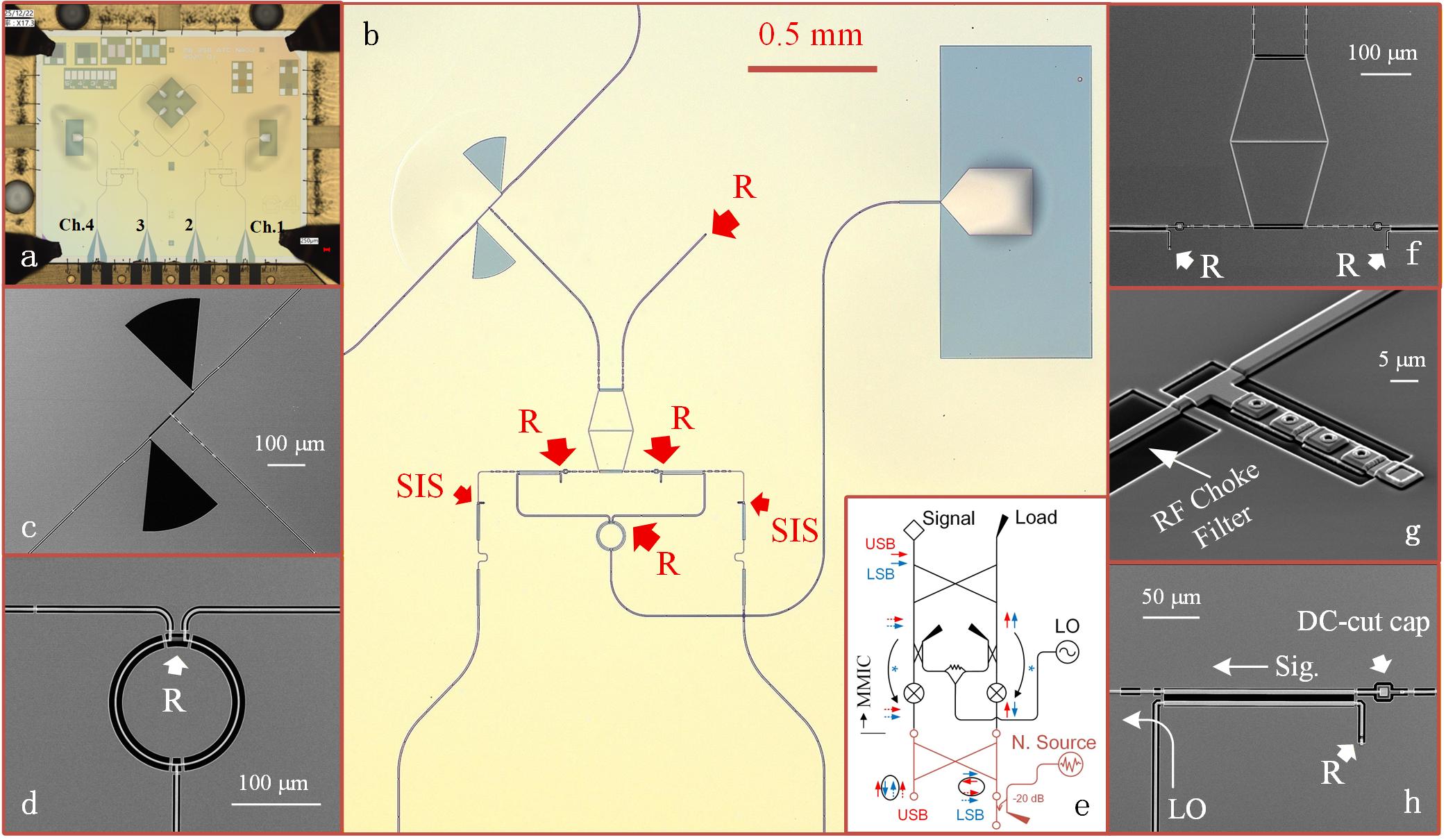}
\caption{Monolithic dual-polarization 2SB Nb SIS MMIC design. (a) Optical photograph of the fabricated MMIC chip. (b) Enlarged view of one polarization branch of the integrated 2SB mixer circuit, where the positions of the SIS junction arrays and thin-film resistors (R) are indicated. The circuit includes the planar OMT output, RF quadrature hybrid coupler, directional couplers for LO injection, Wilkinson LO power divider, SIS mixers, and IF extraction network. (c) SEM image of the anti-phase power combiner used together with the planar OMT probes. (d) SEM image of the Wilkinson LO power divider. (e) Equivalent circuit of the 2SB mixer including the external IF quadrature hybrid. The phase relationships for upper-sideband (USB) and lower-sideband (LSB) signals are illustrated to show the principle of sideband separation. A noise source is weakly coupled to the LSB output port for mixer gain measurement. (f) SEM image of the RF quadrature hybrid coupler. (g) SEM image of the four-junction series SIS array with RF choke filter. (h) SEM image of the directional coupler for LO injection and the DC-block capacitor.}
\label{FigCircuitDesign}
\end{figure*}

\section{Circuit Design}

The 2SB Nb SIS mixer was designed based on the previously demonstrated balanced MMIC architecture \cite{shan2018planar}, while introducing full monolithic 2SB functionality with preserving the same assembly approach and preserving scalability for large-format arrays. Reusing the validated waveguide interface and mixer mount was a key design principle, allowing direct compatibility with the existing balanced-mixer holder and minimizing the engineering effort required for implementation.

The mixer was fabricated on a 13 mm $\times$ 10 mm $\times$ 0.4 mm silicon-on-insulator (SOI) substrate. Fig.~\ref{FigCircuitDesign}(a) shows the overall MMIC chip layout. Four membrane-based probe antennas form the planar orthomode transducer (OMT) at the upper center, while two additional probe antennas couple the LO signal for the two linear polarizations. All probe antennas are fabricated on 6-$\mu$m-thick silicon membranes formed by removing the handle layer from the backside of the SOI substrate at the waveguide intersections.

The membrane probe design and mixer mount are identical to those used in the balanced MMIC mixer \cite{shan2018planar} and are therefore not repeated here. However, preserving the same waveguide interface requires a crossover between the coplanar waveguide (CPW) carrying the LO signal and the CPW for IF extraction in each polarization channel. The membrane-probe performance was independently evaluated using a back-to-back probe-pair structure, and the measured transmission agreed well with theoretical predictions \cite{masukura2023silicon}.

The circuit layout is mirror-symmetric, with each half corresponding to one linear polarization, as shown in Fig.~\ref{FigCircuitDesign}(b), and the equivalent circuit is presented in Fig.~\ref{FigCircuitDesign}(e). All frequency-dependent components were optimized over the RF band of 125--163 GHz. CPW was adopted as the primary transmission-line format because of its lower dissipation loss than microstrip lines and its lower capacitance at IF. The center strip width and gap were designed to be 3 $\mu$m and 4 $\mu$m, respectively, resulting in a characteristic impedance of approximately 64 $\Omega$. Microstrip lines with much lower characteristic impedance, typically around 20 $\Omega$, were selectively used in the RF quadrature hybrid coupler and IF choke filter, where low-impedance sections are required. Intermediate characteristic impedances between 20 $\Omega$ and 64 $\Omega$ were synthesized by combining these two transmission-line formats.

For each polarization, the RF signal collected by a corrugated feed horn is coupled through a pair of OMT probes and combined by an anti-phase power combiner, shown in the SEM image of Fig.~\ref{FigCircuitDesign}(c). The design details of this combiner are described in \cite{shan2018planar}. The combined RF signal then propagates through a quadrature 3-dB branch-line coupler, shown in Fig.~\ref{FigCircuitDesign}(h), which is inherited from the previously demonstrated balanced MMIC mixer \cite{shan2018planar}.

The main modification for 2SB operation is that the isolated port of the quadrature coupler is terminated with a matched thin-film resistive load instead of being left as the LO input port. This enables proper quadrature signal splitting for the two SIS branches required in the 2SB architecture. Each output port is followed by a nominal -15 dB directional coupler for LO injection. The coupling level was chosen as a tradeoff between sufficient LO pumping power and minimal RF insertion loss. The coupler was first designed analytically \cite{wen2003coplanar} and then fine-tuned using full-wave numerical simulation\cite{ansyshfss}. Simulated results show less than 1 dB variation in coupling and input reflection below $-20$ dB across the entire RF band, assuming ideal port matching.

A DC-block capacitor was placed immediately before the directional coupler to electrically isolate the two SIS branches, allowing independent bias control, which is essential for planar 2SB operation. Unlike waveguide structures, superconducting planar transmission lines support DC propagation and therefore do not provide intrinsic electrical isolation between the two mixer branches. The capacitor was designed to be $8\,\mu\mathrm{m} \times 8\,\mu\mathrm{m}$ using a 50-nm-thick $\mathrm{Al_2O_3}$ dielectric layer, corresponding to a capacitance of approximately $0.1\,\mathrm{pF}$, sufficiently large to allow the RF signal passing through. In addition to DC isolation, the capacitor minimizes additional IF capacitance contributed by the upstream circuit, particularly the microstrip sections in the RF quadrature hybrid, thereby helping to preserve broad IF bandwidth.

A four-junction series array, shown in Fig.~\ref{FigCircuitDesign}(g), was adopted to ensure linear response under room-temperature blackbody loading and to broaden the IF bandwidth by reducing the effective junction capacitance. This also improves the dynamic range of the receiver under practical astronomical operating conditions. Each SIS junction has a nominal diameter of $2\,\mu\mathrm{m}$ and a critical current density of approximately $8\,\mathrm{kA/cm^2}$. The far end of the junction array is grounded through a superconducting via, while short CPW sections between adjacent junctions provide both electrical interconnection and tuning inductance. The embedding impedance seen by the junction array was optimized using the in-house simulator SISMA \cite{shan2018sisma}, which is based on the quantum mixing theory of Tucker and Feldman \cite{tucker1985quantum}. The target impedance region was selected to achieve low conversion loss and broad RF bandwidth while maintaining stable mixer operation. A three-section quarter-wavelength RF choke filter was adopted to prevent RF signal leakage into the IF port, with the first high-impedance section shown in Fig.~\ref{FigCircuitDesign}(g).

A Wilkinson power divider was adopted to provide equal-phase LO distribution with good amplitude balance for the two SIS mixers in the 2SB configuration. Although the port isolation of the Wilkinson divider is not critical in the present design because the two mixers are excited through weakly coupled directional couplers, the divider serves as a useful demonstration of integrated LO power distribution in the MMIC platform. Furthermore, the isolation property may become beneficial in future balanced-2SB architectures. Each arm of the divider was designed to be a quarter wavelength long at the band center, with a characteristic impedance of $\sqrt{2}$ times the input CPW impedance, resulting in approximately $90~\Omega$. This was implemented using a CPW geometry of gap/strip/gap = $7\,\mu\mathrm{m}/2\,\mu\mathrm{m}/7\,\mu\mathrm{m}$. The isolation resistor was designed to be twice the input characteristic impedance, approximately $130~\Omega$, following the standard Wilkinson divider condition.

Bridges formed in the base layer were periodically introduced to connect the CPW ground planes on both sides of the center strip and maintain equal electrical potential, thereby suppressing unwanted slotline-mode excitation and ensuring stable CPW propagation. The under-bridges and the crossing center strips were separated by an approximately 300-nm-thick SiO$_2$ insulating layer. Because each bridge is electrically short at the operating frequencies, its individual effect on signal propagation is negligibly small. The bridge spacing was designed to be approximately a quarter wavelength at the band center. This ensures that the first bridge-induced resonance appears well above the operating band and therefore does not affect mixer performance.

The CPW ground planes on the MMIC surface were electrically connected to the mixer holder using bonding wires placed along the chip periphery with a spacing of approximately 1 mm, as shown in Fig.~\ref{FigCircuitDesign}(a). This spacing is sufficiently smaller than the effective wavelength at the IF and therefore provides adequate IF grounding for the CPW structure. No measurable difference in RF performance was observed when the bonding-wire density was further increased within experimental uncertainty. This indicates that bonding-wire grounding is sufficient and does not significantly limit mixer performance in the present frequency range.

\section{Device Fabrication}

The fabrication of the monolithic dual-polarization 2SB SIS MMIC requires the simultaneous integration of membrane-based waveguide probes, superconducting SIS junctions, and thin-film resistors on a single silicon-on-insulator (SOI) substrate while maintaining reasonable yield and stable RF performance. This level of integration is essential for scalable large-format heterodyne arrays but also introduces significant fabrication complexity. Most of the fabrication process follows the previously established balanced SIS mixer MMIC process \cite{ezaki2019fabrication}, which combines a standard Nb/Al/AlO$_x$/Al/Nb SIS multilayer process on the front side of a 3-inch SOI wafer with aligned backside deep etching to form silicon membranes for the waveguide probes. The main process update introduced for the present 2SB mixer is the integration of thin-film resistors for the Wilkinson divider and RF matched terminations. To avoid unnecessary repetition, only this newly introduced resistor process is described in detail here.

\begin{figure}[tb]
\centering
\includegraphics[width=0.40\textwidth,clip]{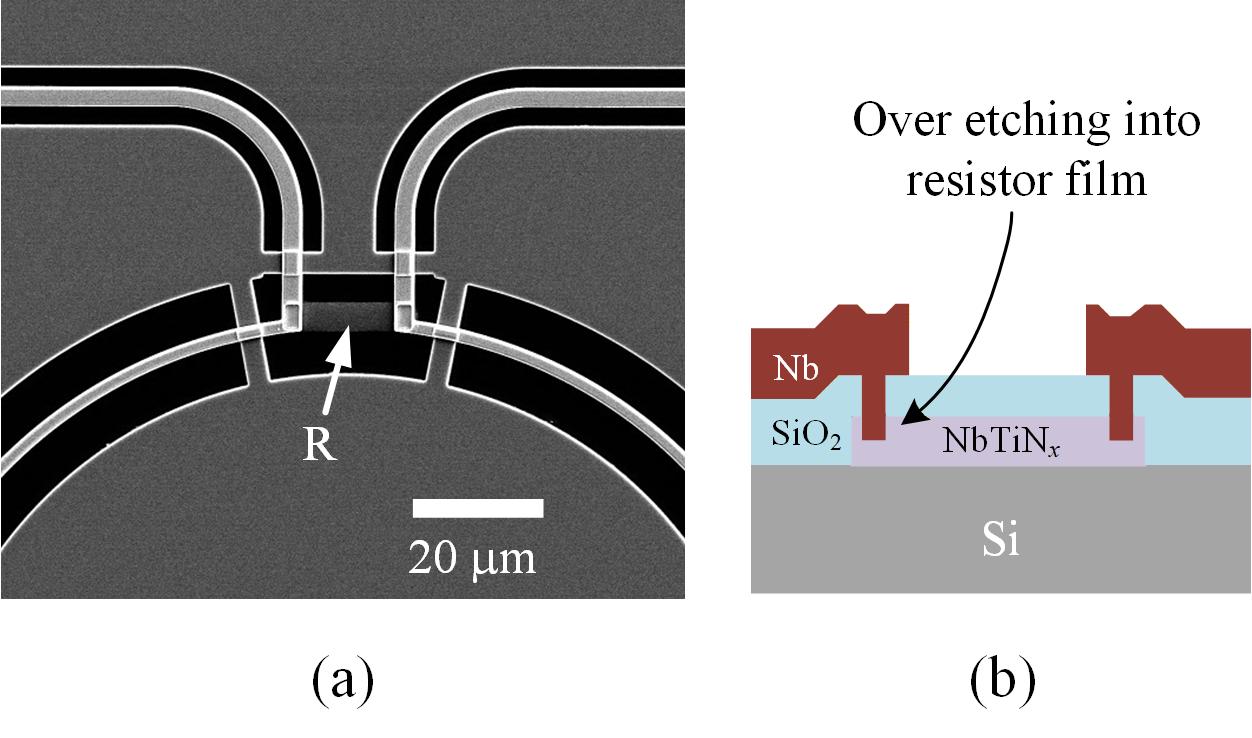}
\caption{Thin-film resistor process for the Wilkinson power divider. (a) SEM image of the lumped-element resistor implemented with an amorphous NbTiN$_x$ film. (b) Schematic cross-sectional view of the resistor structure. The NbTiN$_x$ resistor layer is covered by the SiO$_2$ dielectric and contacted by the Nb wiring layer through via holes. Over-etching during via-hole formation can partially remove the NbTiN$_x$ film, causing significant resistance variation if the film is too thin. This process constraint requires both sufficiently high resistivity and adequate film thickness for stable resistor fabrication.}
\label{FigResistor}
\end{figure}

The lumped-element thin-film resistor used in the Wilkinson divider is shown in the SEM image and schematic layout of Fig.~\ref{FigResistor}. The NbTiN$_x$ film was deposited using N$_2$ reactive sputtering and patterned using a lift-off process. After deposition of the SiO$_2$ layer, via holes were formed by CF$_4$ reactive-ion etching to electrically connect both the resistor pattern and the top electrodes of the SIS junctions to the Nb wiring layer. During this via-hole etching step, the NbTiN$_x$ layer is inevitably exposed to CF$_4$ plasma and suffers additional over-etching. If the resistive film is too thin, this over-etching introduces significant uncertainty in the final resistance value and reduces process reproducibility.

The key fabrication challenge for the thin-film resistor is achieving sufficiently high resistance. The required resistivity of the resistive film is constrained by three factors. First, the resistor width must be larger than $2\,\mu\mathrm{m}$, which is determined by the minimum reproducible line width of the present fabrication process. Second, to minimize parasitic inductance and avoid frequency-dependent impedance mismatch, the resistor length should be shorter than $\lambda/20$, where $\lambda$ is the wavelength in the superconducting thin-film transmission line. For future submillimeter-wave applications up to the niobium gap frequency, this maximum length should be less than $10\,\mu\mathrm{m}$. Together with the minimum width constraint, this requires an aspect ratio of $\eta < 5$. Third, the resistive film must tolerate approximately $50\,\mathrm{nm}$ of over-etching during CF$_4$ via-hole etching, particularly near the wafer edge where cross-wafer etching non-uniformity is significant. This requires a moderate film thickness of approximately $200\,\mathrm{nm}$. Combining these three requirements, a target resistance of $100~\Omega$ requires a film resistivity higher than $400\,\mu\Omega\cdot\mathrm{cm}$. This value is significantly higher than that of conventional metallic resistor films such as NiCr, Al alloys, Pd, and Mo, which are typically below $10\,\mu\Omega\cdot\mathrm{cm}$, and also higher than most reported compound resistive films such as NbO, NbN, MoN, and TiN, which are typically below $200\,\mu\Omega\cdot\mathrm{cm}$. Conventional resistor materials are therefore fundamentally unsuitable for this application.

Amorphous NbTiN$_x$ was selected for the thin-film resistors because it simultaneously provides sufficiently high normal-state resistivity and full compatibility with the existing SIS fabrication resource, including the available NbTi target originally used for high-gap superconducting films in THz SIS mixers. The superconducting transition temperature ($T_c$) of NbTiN$_x$ strongly depends on both the nitrogen content and the film texture. By controlling sputtering parameters such as nitrogen flow rate, total pressure, and sputtering current, $T_c$ can be intentionally reduced below the operating temperature of the SIS mixer, allowing the film to function as a normal resistive element at 4 K. To enable systematic optimization in this multidimensional process space, we established a theoretical model relating the nitrogen content $x$ to the current-voltage characteristics (IVCs) of the magnetron discharge, which provides a convenient in-situ process observable, and further to the sputtering conditions \cite{shan2021modeling}. Guided by this model, the sputtering conditions were optimized to achieve a resistivity of approximately $800\,\mu\Omega\cdot\mathrm{cm}$ with $T_c < 3.7~\mathrm{K}$ \cite{shan2022cryogenic}.

The electrical properties of the NbTiN$_x$ resistor film were verified using a test pattern located at the upper right corner of each MMIC chip. The sheet resistance was measured by the Van der Pauw method \cite{van1991method}, allowing direct evaluation of the resistor properties on the fabricated device wafer. For films with a resistivity of approximately $800\,\mu\Omega\cdot\mathrm{cm}$, the resistance typically increases by about 10\% when cooled from room temperature to 4 K. The nitrogen composition was measured by energy-dispersive X-ray spectroscopy (EDS) to be approximately $x \approx 0.3$, indicating that the films are significantly nitrogen-deficient, consistent with the low-$T_c$ high-resistivity condition predicted by the sputtering model \cite{shan2021modeling}.

Another important process modification was the removal of the 100-nm Al$_2$O$_3$ etch-stop layer previously deposited before the Nb/Al/AlO$_x$/Al/Nb multilayer (the base layer) deposition. This amorphous layer had been introduced to protect the wafer from over-etching during the formation of the SIS base electrode pattern through full base layer etching. However, we found that this additional dielectric layer could significantly hinder heat diffusion from the SIS junctions to the silicon substrate and thereby increase junction self-heating \cite{shan2026self}. Since excessive self-heating can degrade mixer performance by modifying the junction I–V characteristics and reducing conversion efficiency, minimizing this thermal resistance is important for stable SIS operation. Although the quantitative impact of this self-heating on RF performance has not yet been fully established, the Al$_2$O$_3$ etch-stop layer was removed as a preventive measure to improve thermal coupling and reduce potential performance degradation.

\section{Measurement Setup}

\begin{figure}[tb]
\centering
\includegraphics[width=0.40\textwidth,clip]{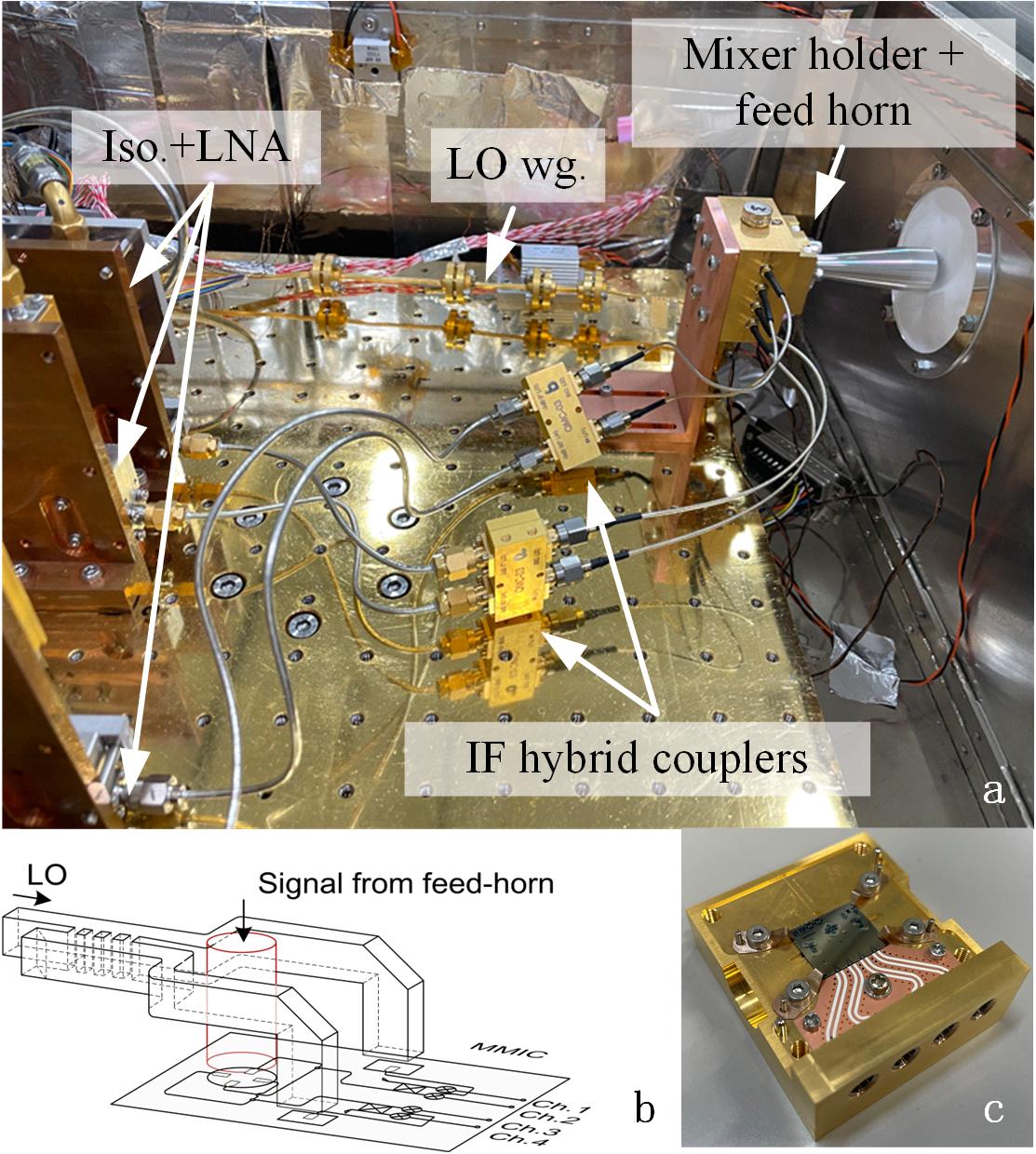}
\caption{Measurement setup for evaluating the monolithic dual-polarization 2SB MMIC SIS mixer. (a) Photograph of the 4-K stage inside the cryostat, showing the mixer holder with feed horn, LO waveguide input, cryogenic IF hybrid couplers, isolators, and LNAs. The IF hybrid couplers are placed outside the mixer holder to complete the 2SB configuration. (b) Schematic illustration of the internal structure of the mixer holder. The LO signal is divided inside the holder by a waveguide 3-dB branch-line coupler and coupled to the MMIC through membrane-based probes, while the RF signal is injected through the corrugated feed horn. Although only a single pixel is measured, this structure preserves the key HPI architecture scalable to large-format arrays. (c) Photograph of the bottom plate of the mixer holder with the MMIC chip mounted and the four IF output channels connected through the PCB interface.}
\label{FigSetup}
\end{figure}

 The 2SB MMIC mixer chip was evaluated in a 4-K cryostat cooled by a Gifford–McMahon cooler, as shown in Fig.~\ref{FigSetup}(a). The mixer holder and corrugated feed horn assembly were aligned to the center of the vacuum window, which consisted of a 25-$\mu$m-thick Kapton film at room temperature and a 100-$\mu$m-thick ZITEX filter at approximately 30 K. The RF input noise contributed by the optical path was independently measured to be 5–7 K across the RF band using a balanced MMIC mixer \cite{shan2023inconstant}. The LO signal was generated using a GUNN oscillator and coupled into the mixer holder through a waveguide. A GUNN oscillator was selected because of its low noise. In contrast, alternative LO sources in the laboratory based on a signal generator and multiplier chain exhibited pronounced broadband phase noise and AM spurious. Two permanent magnets were placed outside the mixer holder to suppress the Josephson current and minimize Josephson noise in the SIS mixers. One of the magnets is visible in Fig.~\ref{FigSetup}(a), attached to the top of the holder.

Although only a single pixel was measured, the mixer holder preserves all key architectural features of the HPI concept scalable to large-format arrays, as illustrated in Fig.~\ref{FigSetup}(b). In particular, the LO distribution network and the MMIC mounting scheme are identical to those required for array implementation. The same holder architecture was previously used for the balanced MMIC mixer \cite{shan2018planar}. The LO signal is divided inside the holder using a 3-dB branch-line directional coupler, with the two output waveguides bending downward in the H-plane and coupling to the MMIC chip through on-chip membrane probes. A branch-line coupler was adopted instead of a simpler Y-junction power divider because of its higher isolation between the two output ports. This isolation is important for minimizing LO-path crosstalk, which would otherwise degrade cross-polarization performance \cite{shan2019experimental}. A magic-T-based LO divider was also tested and showed superior amplitude balance and bandwidth \cite{Douglas2026Preparation}. Fig.~\ref{FigSetup}(c) shows the bottom plate of the holder with the MMIC chip mounted. The four IF output channels were routed through a printed circuit board (PCB) board to feed-through SMP coaxial connectors located on the sidewall of the holder.

External to the mixer holder, two cryogenic IF hybrid couplers (QMC-03) complete the 2SB configuration by recombining the two IF branches. Since amplitude and phase imbalance in the IF hybrids can directly limit the measured sideband rejection ratio, the couplers were independently characterized in advance to confirm that their imbalance was sufficiently small and therefore did not dominate the measurement. The outputs from the IF hybrid couplers were connected to 4–8 GHz band isolators followed by low-noise amplifiers (LNAs), all mounted at the 4-K stage. The LNA for Ch.~1 (Low Noise Factory LNF-LNC 0.3–14 GHz) differs from those used for the other channels (Nitsuki MODEL9891, 4–8 GHz) in noise temperature. The former provides a measured noise temperature of approximately 5 K across 4–8 GHz, while the latter provides approximately 10 K, both including the isolators.

The receiver noise temperature was measured using the standard Y-factor method with hot and cold loads at room temperature (295 K) and liquid-nitrogen temperature (77 K), respectively. The reported values are single-sideband (SSB) receiver noise temperatures derived from the Y-factor measurement after correction for the finite sideband rejection \cite{kerr2001alma}.

The sideband ratio was determined by measuring the IF responses to a weak continuous-wave (CW) signal injected through the feed horn together with the hot load \cite{kerr2001alma}. The CW power was carefully adjusted to remain sufficiently low so that the SIS mixers operated in the linear regime without gain compression or saturation. By comparing the responses at the upper and lower sidebands, the intrinsic image rejection of the MMIC was evaluated.

\section{RF Performance}

The RF performance of the 2SB MMIC was characterized in terms of receiver noise temperature and sideband rejection ratio. Since low receiver noise is the primary requirement for heterodyne receivers, the receiver noise temperature is first evaluated. After confirming acceptable sensitivity, the effectiveness of sideband separation is assessed through the measured sideband ratio.

\begin{figure}[tb]
\centering
\includegraphics[width=0.40\textwidth,clip]{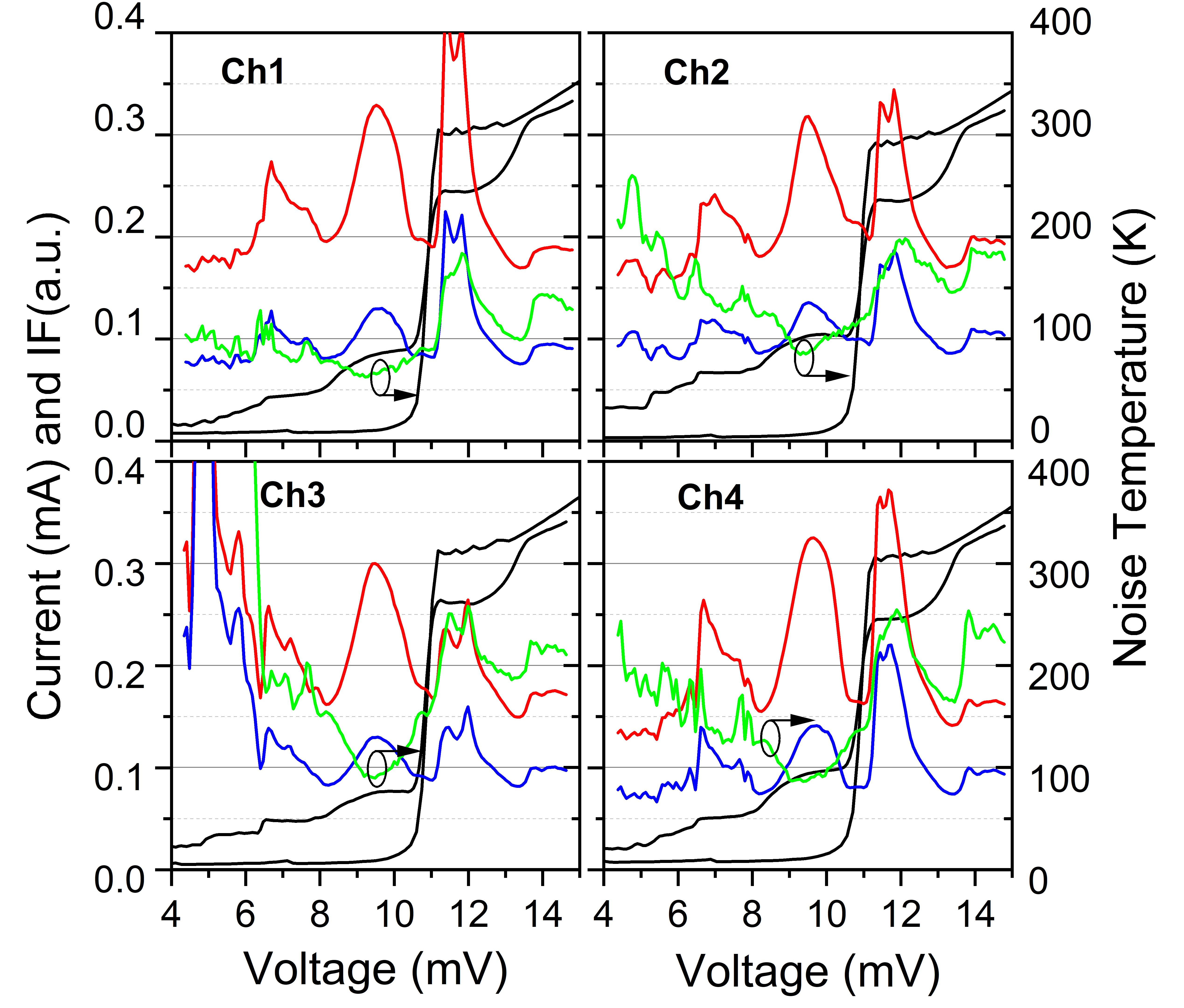}
\caption{Pumped and unpumped I–V curves and IF responses of the four channels measured at $f_{\mathrm{LO}} = 153$ GHz. Channel definitions are given in Fig.~\ref{FigCircuitDesign}(a). The black curves show the pumped and unpumped SIS I–V characteristics. The red and blue curves represent the IF output power at 6 GHz for hot-load and cold-load measurements, respectively. The green curves show the corresponding uncorrected receiver noise temperature. During the bias sweep of each channel, the other mixer in the same 2SB unit was fixed near the center of the first photon-assisted step below the gap voltage.}
\label{Fig153GHz}
\end{figure}

The mixer response to thermal radiation input and the corresponding noise temperature measured at 153 GHz are shown as functions of bias voltage in Fig.~\ref{Fig153GHz}. The two SIS mixers forming one 2SB unit exhibited a pumping-height difference of approximately 20\% (0.8 dB), indicating reasonably balanced LO injection. Using the same method, the LO power imbalance was measured to be less than 1 dB across the RF band, confirming acceptable performance of the on-chip Wilkinson power divider. The hot and cold responses were measured by sweeping the bias voltage of one SIS mixer in the 2SB unit while keeping the other fixed at the center of the first photon-assisted step below the gap voltage, where stable LO pumping was obtained.

\begin{figure}[tb]
\centering
  \begin{subfigure}{0.4\textwidth}
    \centering
    \includegraphics[width=0.8\textwidth,clip]{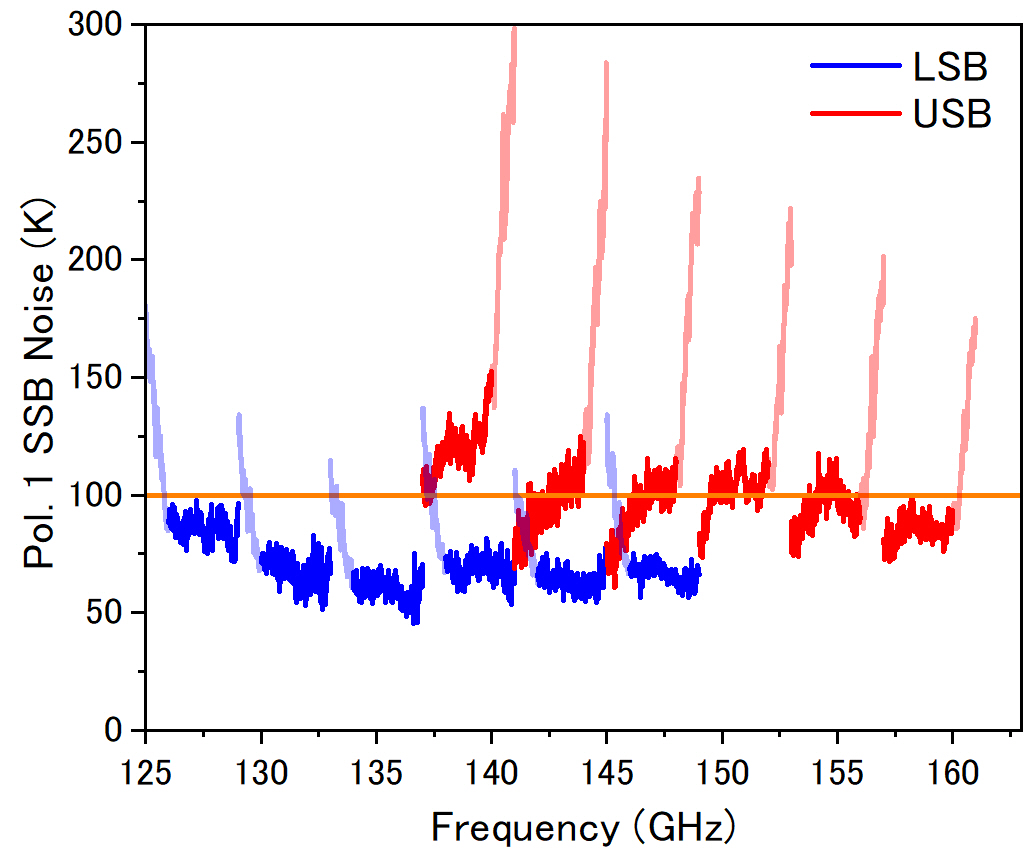}
    \caption{ }
  \end{subfigure}
  % second figure-------------------------
  \begin{subfigure}{0.4\textwidth}
    \centering
    \includegraphics[width=0.8\textwidth,clip]{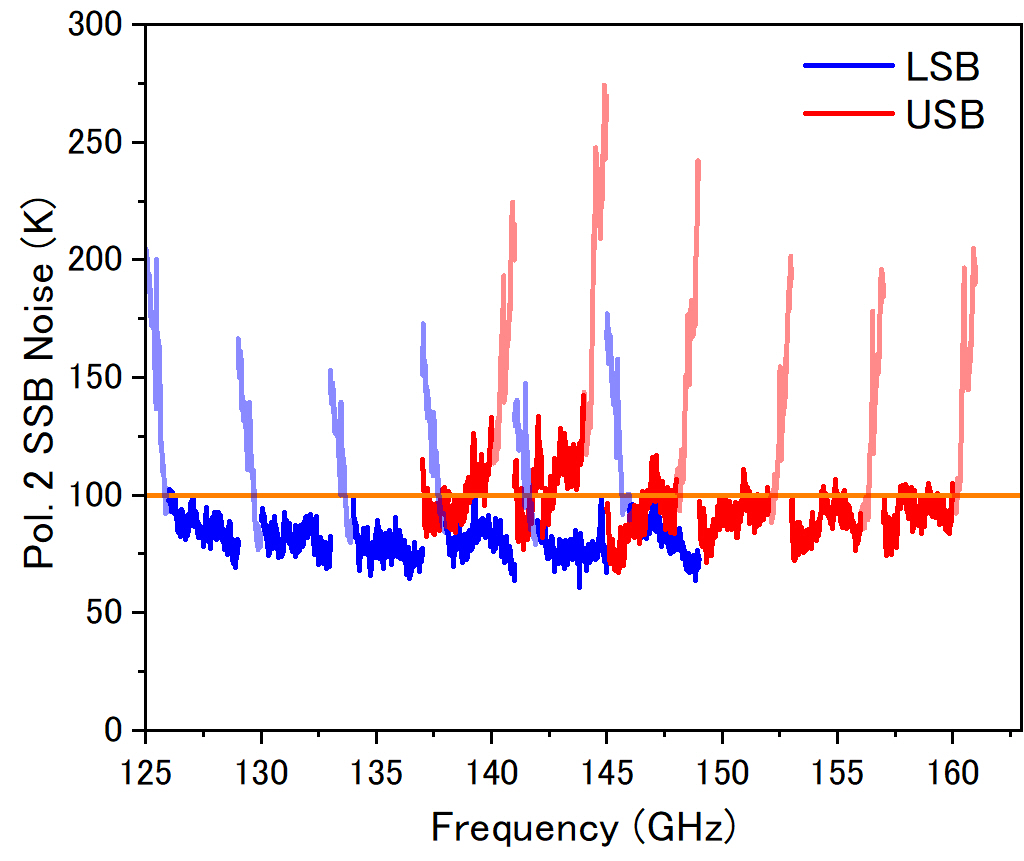}
    \caption{ }
  \end{subfigure}

\caption{SSB receiver noise temperature corrected for the finite sideband ratio, measured over the full RF band for polarization 1 (a) and polarization 2 (b). The blue and red curves represent the LSB and USB, respectively. A minimum SSB noise temperature of approximately 60 K was achieved in polarization 1 LSB, where the corresponding IF chain used the lowest-noise cryogenic LNA. The other channels showed minimum noise temperatures of approximately 80 K. The orange horizontal line at 100 K indicates a practical reference level for receiver performance. Similar noise performance was obtained for both polarizations.}
\label{FigNoise}
\end{figure}

The receiver noise temperature measured using the Y-factor method and corrected for the finite sideband ratio \cite{kerr2001alma} is shown in Fig.~\ref{FigNoise} across the full RF band. The LO frequency was tuned in six steps of 4 GHz to cover the operating range. The SSB receiver noise temperature varies from approximately 100 K near the band edges to a minimum of about 60 K in Ch.~1, while approximately 80 K was achieved in the other channels. These values indicate that the monolithic 2SB MMIC preserves practical receiver sensitivity for astronomical observations despite the increased circuit complexity introduced by full planar integration. The lower noise temperature observed in Ch. 1 is attributed to the lower noise of its IF LNA rather than to intrinsic mixer performance. This conclusion was confirmed by reversing the LNA connections, after which Ch. 4 became the lowest-noise channel.

The minimum SSB receiver noise temperature measured in Ch.~1, including the OMT contribution, was approximately 60 K. This value is already competitive for practical astronomical use, considering that the ALMA receiver specification at this frequency was 82 K \cite{ALMAReceiverNoise}. However, it remains higher than the state-of-the-art conventional waveguide SIS mixers operating in the same frequency band, which achieve approximately 40 K \cite{asayama2014development}. A major source of noise that is specific to MMICs of this excess noise is the use of electrically long superconducting thin-film transmission lines, which introduce higher dissipation loss than the hollow metal waveguides used in conventional waveguide SIS mixers. To quantify this effect, we measured the losses of coplanar waveguides (CPWs) and microstrip lines using half-wavelength resonators and separated the total dissipation into conductive loss, dielectric loss, radiation loss, and flux-flow loss components \cite{shan2024separation}. The measured losses were approximately 0.09 dB per wavelength for CPWs and 0.25 dB per wavelength for microstrip lines at this frequency band. The total transmission loss between the OMT and the SIS junctions was estimated to be approximately 0.7 dB, corresponding to approximately four wavelengths of CPW and one wavelength of microstrip line. This is larger than the insertion loss of conventional metal OMTs (0.3–0.5 dB measured at room temperature \cite{2008stt..conf..244A}, and further reduced by approximately 60\% at 4 K \cite{finger2008microwave}). This additional transmission loss accounts for an increase of approximately 7 K in receiver noise temperature. Therefore, transmission-line dissipation is an important but not dominant contributor to the remaining noise gap between the MMIC and conventional waveguide SIS mixers.

Another important contributor to the higher receiver noise is the lower conversion gain of the MMIC mixer itself. To quantify this effect, we measured the conversion gains of the MMIC 2SB mixer and a conventional ALMA Band 4 2SB mixer using a calibrated microwave noise source weakly coupled (approximately -20 dB) into the IF chain at the input of the cryogenic isolators, as shown in Fig. \ref{FigCircuitDesign}e. Because the coupling is weak, the additional noise from the source does not significantly affect the receiver noise temperature when the source is turned off. When the source is turned on and the SIS mixer is biased near the center of the gap voltage, the IF impedance of the SIS mixer, which is approximately given by the dynamic resistance, becomes much smaller than the IF chain impedance. As a result, the intrinsic mixer noise is strongly suppressed at the IF output due to severe impedance mismatch, and the injected IF noise source becomes the dominant contribution at the IF input. Under this condition, the calibrated source serves as an in-situ reference for determining the effective mixer conversion gain. Using this method, the measured conversion gains at 6 GHz were approximately -1 dB for the ALMA Band 4 mixer and  -6 dB for the MMIC mixer. This approximately 5 dB difference increases the effective contribution of the IF chain noise (approximately 5 K for Ch.~1) when referred to the receiver input, resulting in an additional noise penalty of approximately 16 K.

These two factors, transmission-line dissipation and lower mixer conversion gain, reasonably account for the higher receiver noise observed in the MMIC 2SB mixer compared with conventional waveguide SIS mixers. The transmission loss can be further reduced from approximately 0.7 dB to 0.4 dB by removing the dielectric layer under the CPW center strips to eliminate unnecessary dielectric loss and by avoiding magnetic fields perpendicular to the MMIC chip to minimize flux-flow loss \cite{shan2024separation}. This improvement alone would reduce the receiver noise penalty associated with planar integration. The origin of the considerably lower conversion gain of the MMIC 2SB SIS mixer is not yet fully understood. If it mainly arises from unintended embedding-impedance offsets, it can likely be improved by redesigning the mixer circuit and further improving fabrication accuracy. This will be the subject of future work.

A systematic increase in receiver noise temperature was also observed when the IF exceeded approximately 7 GHz. As shown in Fig.~5, this appears as rising tails at the lower-frequency edge of the lower sideband (LSB) and the higher-frequency edge of the upper sideband (USB). This behavior was consistently observed in all measured 2SB MMIC mixers, indicating that it is a systematic IF-related issue rather than device-to-device variation. It is highly likely that the origin is inside the MMIC itself rather than in the external IF chain, because the same external IF system was used for evaluating the balanced MMIC mixers, where no comparable high-IF noise increase was observed. Clarifying this issue will be an important subject of future work.

Gain compression was measured at several RF frequencies using the standard method of injecting a weak CW signal through the LO port under both cold-load and hot-load background conditions \cite{kerr2003measurement}. The measured gain compression was essentially zero within an experimental uncertainty of 1\%, indicating that the mixer operated in a sufficiently linear regime under normal thermal loading and that the reported receiver noise temperature was not affected by saturation.

\begin{figure}[tb]
\centering
  \begin{subfigure}{0.4\textwidth}
    \centering
    \includegraphics[width=0.8\textwidth,clip]{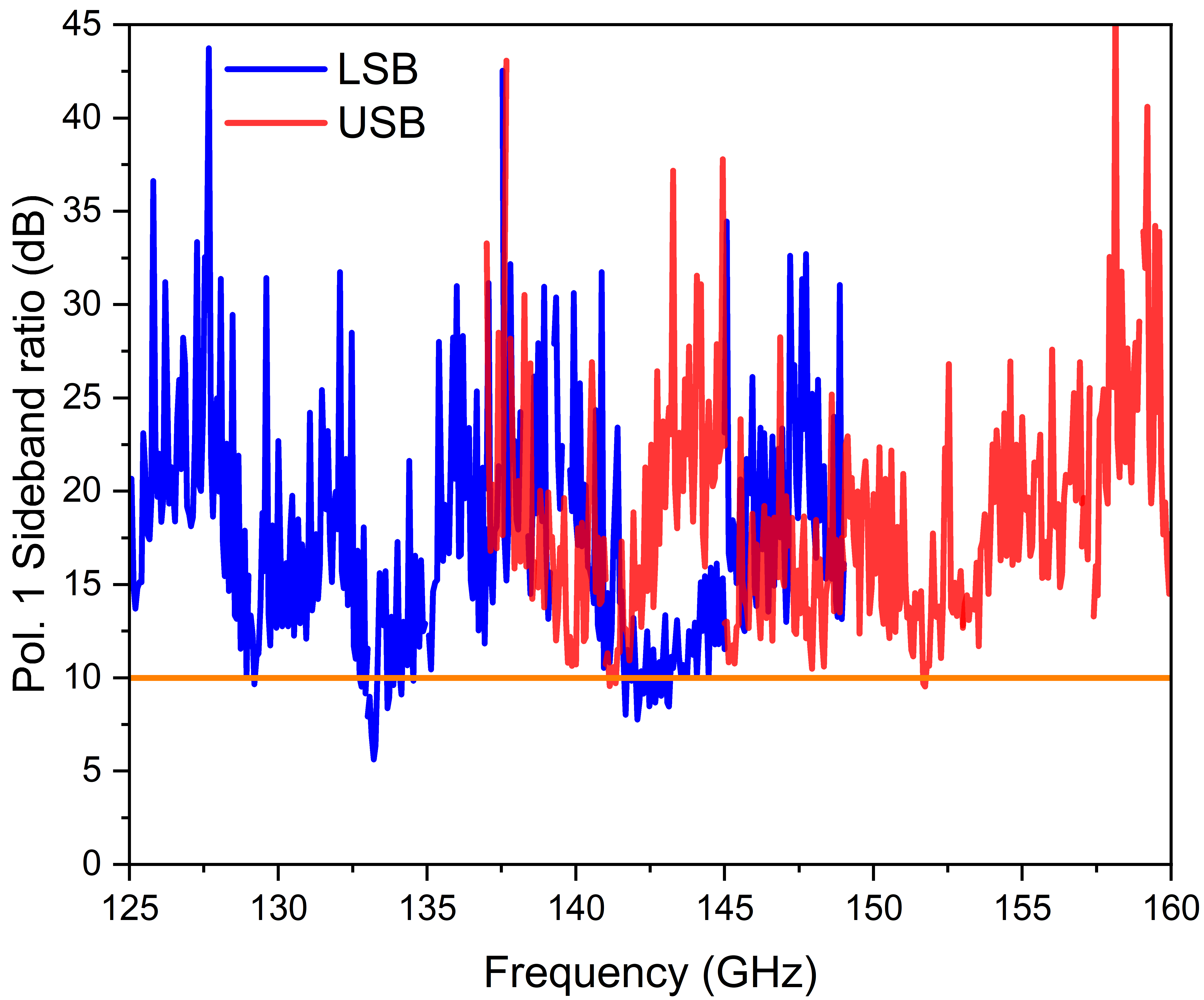}
    \caption{ }
  \end{subfigure}
  % second figure-------------------------
  \begin{subfigure}{0.4\textwidth}
    \centering
    \includegraphics[width=0.8\textwidth,clip]{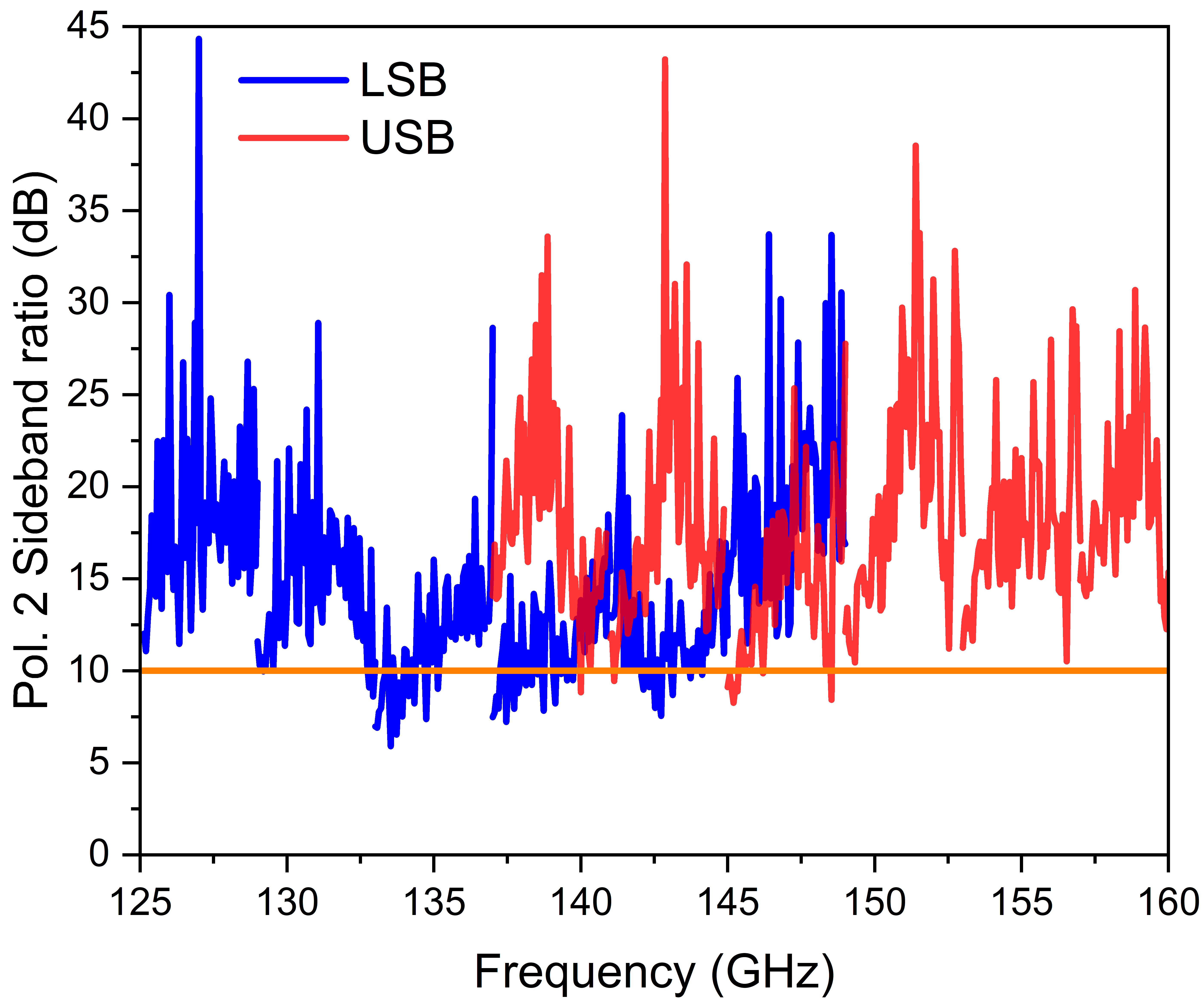}
    \caption{ }
  \end{subfigure}
\caption{Measured SBR of polarization 1 (a) and polarization 2 (b) over the full RF band. The blue and red curves represent the LSB and USB, respectively. The orange horizontal line at 10 dB indicates the minimum practical requirement for effective image-sideband suppression in ground-based astronomical observations. Except for a small portion of the RF band, the measured SBR exceeds 10 dB, demonstrating effective sideband separation and confirming the practical functionality of the fully monolithic 2SB MMIC architecture. Similar performance was obtained for both polarizations.}
\label{FigSBR}
\end{figure}

For practical astronomical observations, both low receiver noise and sufficient image-sideband suppression are required. In particular, a sideband rejection ratio (SBR) above approximately 10 dB is generally considered the minimum practical requirement for reducing atmospheric noise contribution from the image sideband in ground-based observations. The measured SBR over consecutive IF bands corresponding to six LO tuning frequencies is shown in Fig.~\ref{FigSBR}. Similar performance was obtained for both linear polarizations. Except for a small portion of the RF band, the measured SBR exceeds 10 dB, demonstrating effective sideband separation and confirming the practical functionality of the fully monolithic 2SB MMIC architecture. Achieving an SBR at the 10-dB level had been difficult for several years because of IF crosstalk, which is a unique issue in planar MMIC implementations. This issue is discussed in detail in the next section.

The near-field co-polarization and cross-polarization beam patterns were measured using a CW source mounted on an external X–Y–$\theta$ scanner. Since the planar OMT design is identical to that used in the previously demonstrated balanced MMIC SIS mixers, similar polarization performance was expected. The measured cross-polarization level was approximately -20 dB, consistent with the balanced MMIC result \cite{shan2019experimental}. This confirms that the additional circuit complexity introduced by the monolithic 2SB architecture does not significantly degrade polarization purity.

\section{IF Crosstalk and Solution}

Achieving practical sideband rejection in monolithic 2SB MMIC mixers was prevented for several years by unexpected IF crosstalk between mixer branches. This issue is unique to highly integrated planar MMIC implementations and does not appear in conventional waveguide SIS mixers. Identifying its physical origin and developing a scalable mitigation strategy were therefore essential for validating the feasibility of the HPI architecture for large-format heterodyne arrays.

\begin{figure}[tb]
\centering
\includegraphics[width=0.35\textwidth,clip]{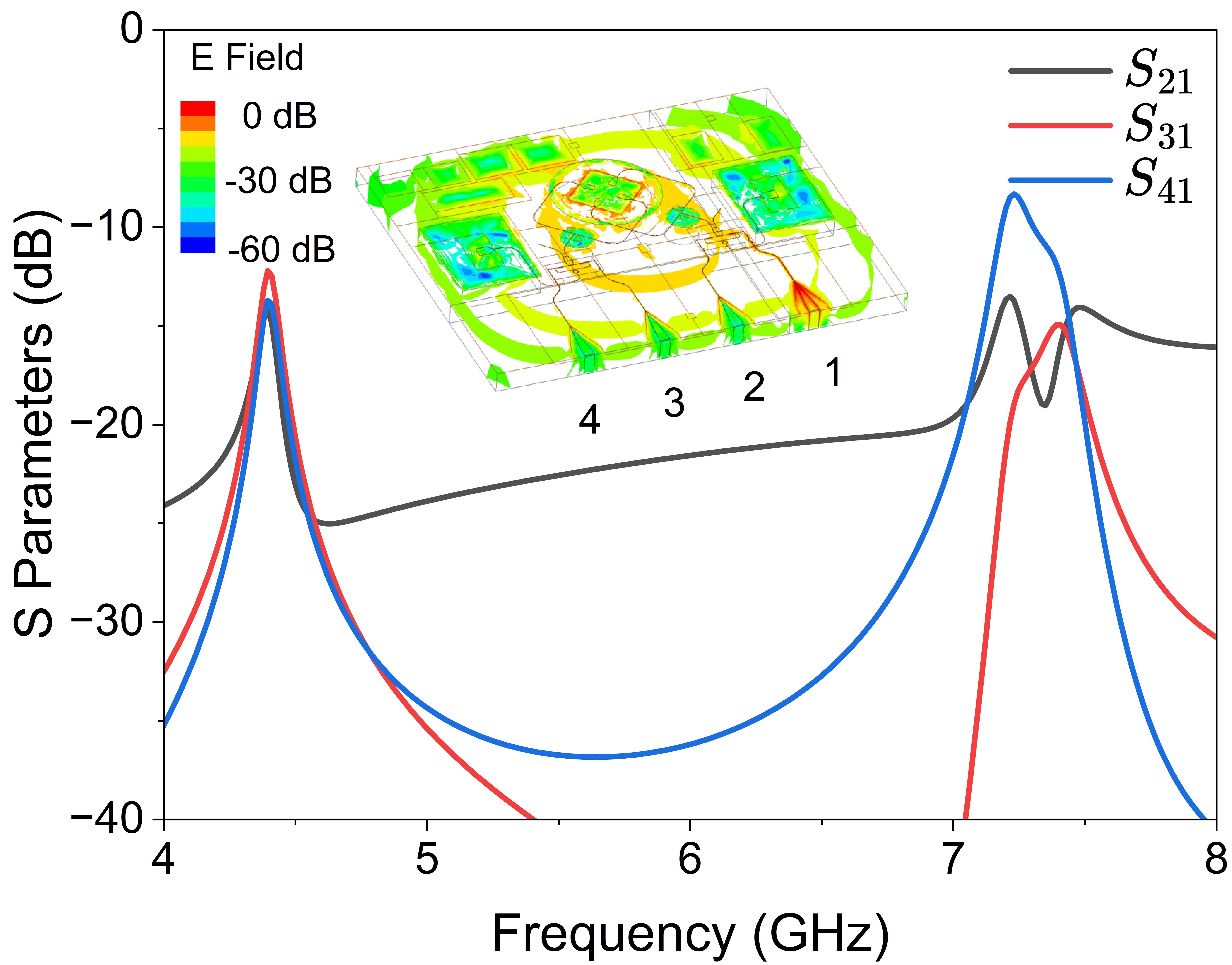}
\caption{HFSS simulation of inter-channel IF crosstalk caused by substrate cavity resonances. The transmission coefficients from Ch.~1 to the other three channels ($S_{21}$, $S_{31}$, and $S_{41}$) show pronounced resonance peaks at approximately 4.4, 7.1, and 7.6 GHz. These resonance frequencies are consistent with the measured degradation of sideband rejection ratio. The inset shows the electric-field distribution of the first resonance mode at 4.4 GHz, confirming that the silicon substrate behaves as a dielectric-filled cavity. }
\label{FigIFXTalk3}
\end{figure}

The IF crosstalk problem was first identified from narrow-frequency dips in the measured SBR, where values below 10 dB were repeatedly observed near fixed IF frequencies of approximately 4.7 GHz and 7 GHz during early RF evaluation. Because these dips always appeared at nearly fixed IF frequencies rather than fixed RF frequencies, substrate-related resonance at IF was strongly suspected. This led us to consider cavity modes formed inside the substrate. Since most of the MMIC surface is covered by superconducting Nb film serving as the ground plane of the coplanar waveguide (CPW), and the substrate is spring-pressed to the copper holder and wire-grounded at the periphery, the silicon substrate effectively behaves as a dielectric-filled cavity. Its resonance frequencies can be approximately estimated by
\begin{equation}\label{EqCavity}
  f_{mn}=\frac{c}{2\sqrt{\varepsilon_r}}\sqrt{\left(\frac{m}{a}\right)^2+\left(\frac{n}{b}\right)^2},
\end{equation}
where $a=\rm{13\,mm}$ and $b=\rm{10\,mm}$ are the length and width of the MMIC chip, respectively; $c$ is the speed of light in vacuum; $m$ and $n$ are positive integers; and $\varepsilon_r=12$ is the relative dielectric constant of silicon. The estimated resonance frequencies of modes (1,1), (1,2), and (2,1) are approximately 5.8 GHz, 6.9 GHz, and 7.2 GHz, respectively, all falling within the IF band and close to the observed SBR degradation frequencies. Although this simple estimation is not exact because the substrate is partially removed beneath the membrane areas, it strongly suggests that substrate cavity modes are responsible for the IF crosstalk. To evaluate this more realistically, full-wave electromagnetic simulation was performed using HFSS \cite{ansyshfss}. The simulated transmission coefficients from Ch.~1 to the other three channels are shown in Fig.~\ref{FigIFXTalk3}, and the inset shows the electric-field distribution of the first resonance mode (1,1).

The ultimate solution for eliminating these substrate resonance modes is the introduction of conductive through-substrate vias (TSVs), for which we have been developing a dedicated fabrication process \cite{ezaki2023development}. By electrically connecting the top Nb ground plane to the metal holder through the silicon substrate, TSVs can effectively suppress cavity-mode formation. However, implementing TSVs in the present MMIC platform is considerably more challenging than in conventional silicon interconnect technologies because the process must remain fully compatible with the backside membrane fabrication used for the waveguide probe antennas. Since several fabrication issues remain unresolved before this TSV process can be reliably applied to working mixer chips, an alternative interim solution was pursued to enable practical demonstration of the monolithic 2SB MMIC before full TSV integration becomes available.

\begin{figure}[tb]
\centering
\includegraphics[width=0.40\textwidth,clip]{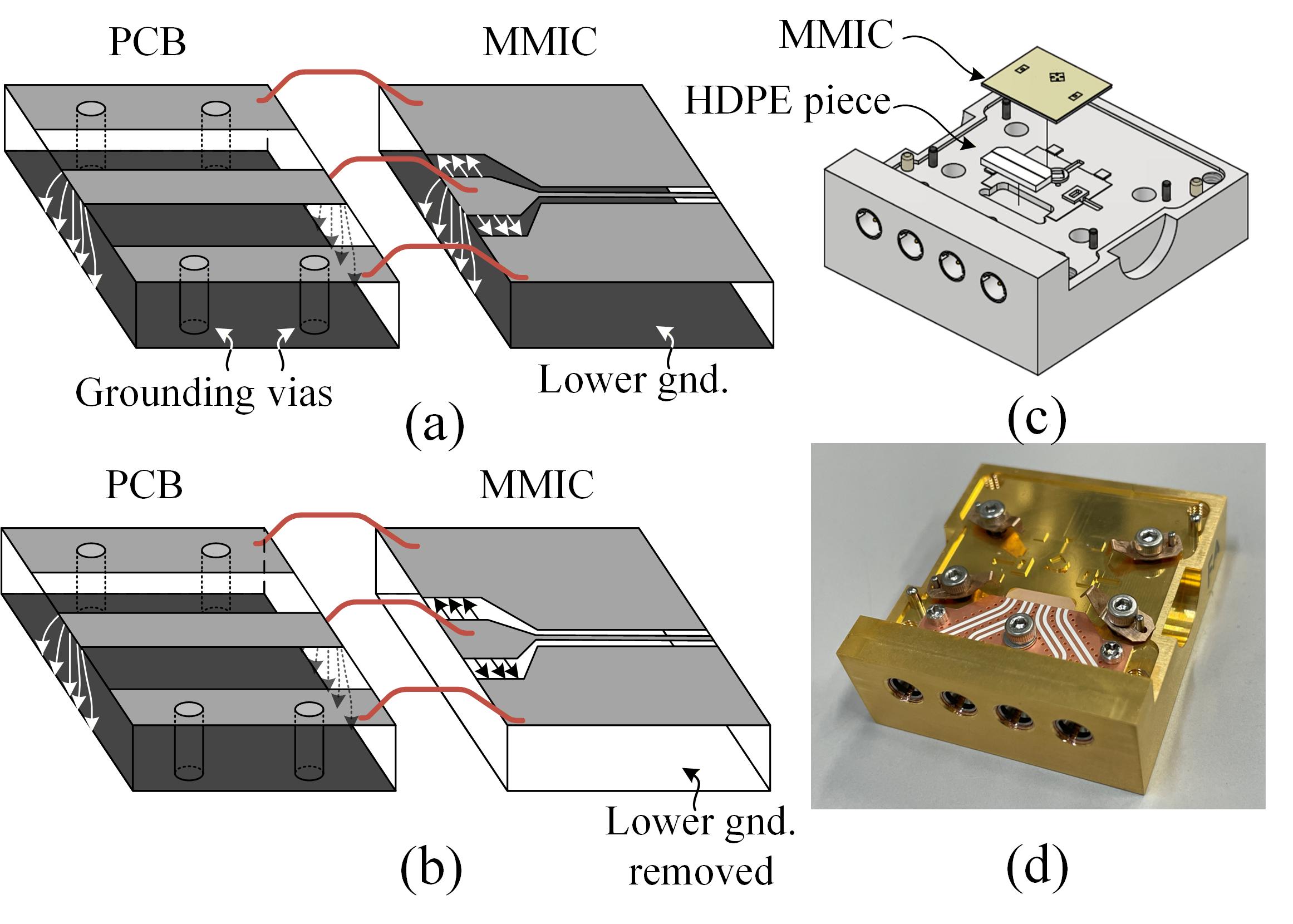}

\caption{Mitigation of IF crosstalk by suppressing the microstrip-mode component at the MMIC–PCB interface. (a) Schematic illustration of the original structure, where the grounded CPW on the MMIC side partially couples to a microstrip-like mode through the lower ground plane beneath the chip. This mode excites substrate cavity resonances and increases inter-channel IF crosstalk. The red lines indicate bonding wires connecting the MMIC and PCB. (b) Modified structure after partially removing the lower ground plane beneath the MMIC. The reduced capacitance under the strip suppresses the microstrip-mode component and weakens coupling to substrate cavity modes, while preserving the CPW mode required for IF transmission. (c) CAD exploded view of the modified mixer holder, where material beneath the MMIC–PCB boundary is removed and replaced with HDPE insert as a mechanical support for ultrasonic wire bonding. (d) Photograph of the modified bottom plate of the mixer holder before mounting the MMIC chip.}

\label{FigFieldMatch}
\end{figure}

The alternative solution was derived from the electric-field distribution at the MMIC–PCB interface, as shown in Fig.~\ref{FigFieldMatch}(a). The PCB is a Rogers RO4360 laminate with relative dielectric constant $\varepsilon_r \sim 6.15$ and thickness of 0.305 mm. The strip width and gap are 0.38 mm and 0.5 mm, respectively. Although the PCB transmission line appears geometrically similar to a grounded CPW, its dominant propagation mode is effectively microstrip-like, with the electric field primarily concentrated beneath the conducting strip. In contrast, the widened bonding pad region on the MMIC side is a grounded CPW supporting both a conventional CPW mode, with electric field mainly in the gap, and an additional microstrip-like mode, with electric field extending into the substrate beneath the strip. Only the microstrip-like mode efficiently couples to the substrate cavity modes, because its electric field is parallel to the cavity-mode field distribution inside the silicon substrate. The CPW mode, whose electric field is mainly perpendicular to the substrate cavity field, contributes little to this coupling. Therefore, suppressing the microstrip-mode component on the MMIC side reduces the excitation of substrate cavity resonances and consequently lowers IF crosstalk. This can be achieved by partially removing the metal ground beneath the MMIC chip, as illustrated in Fig.~\ref{FigFieldMatch}(b). Based on this consideration and verified by electromagnetic simulation, the mixer holder was modified by removing approximately 1 mm depth of material beneath the MMIC and filling the void with a piece of high-density polyethylene (HDPE), as shown in the CAD drawing in Fig.~\ref{FigFieldMatch}(c) and the photograph in Fig.~\ref{FigFieldMatch}(d). The HDPE insert provides sufficient mechanical support for ultrasonic wire bonding of the IF/DC connections. Although the bottom ground plane is not completely removed, the reduced capacitance beneath the strip effectively suppresses the microstrip-mode component and weakens cavity-mode excitation.

\begin{figure}[tb]
\centering
  \begin{subfigure}{0.4\textwidth}
    \centering
    \includegraphics[width=0.8\textwidth,clip]{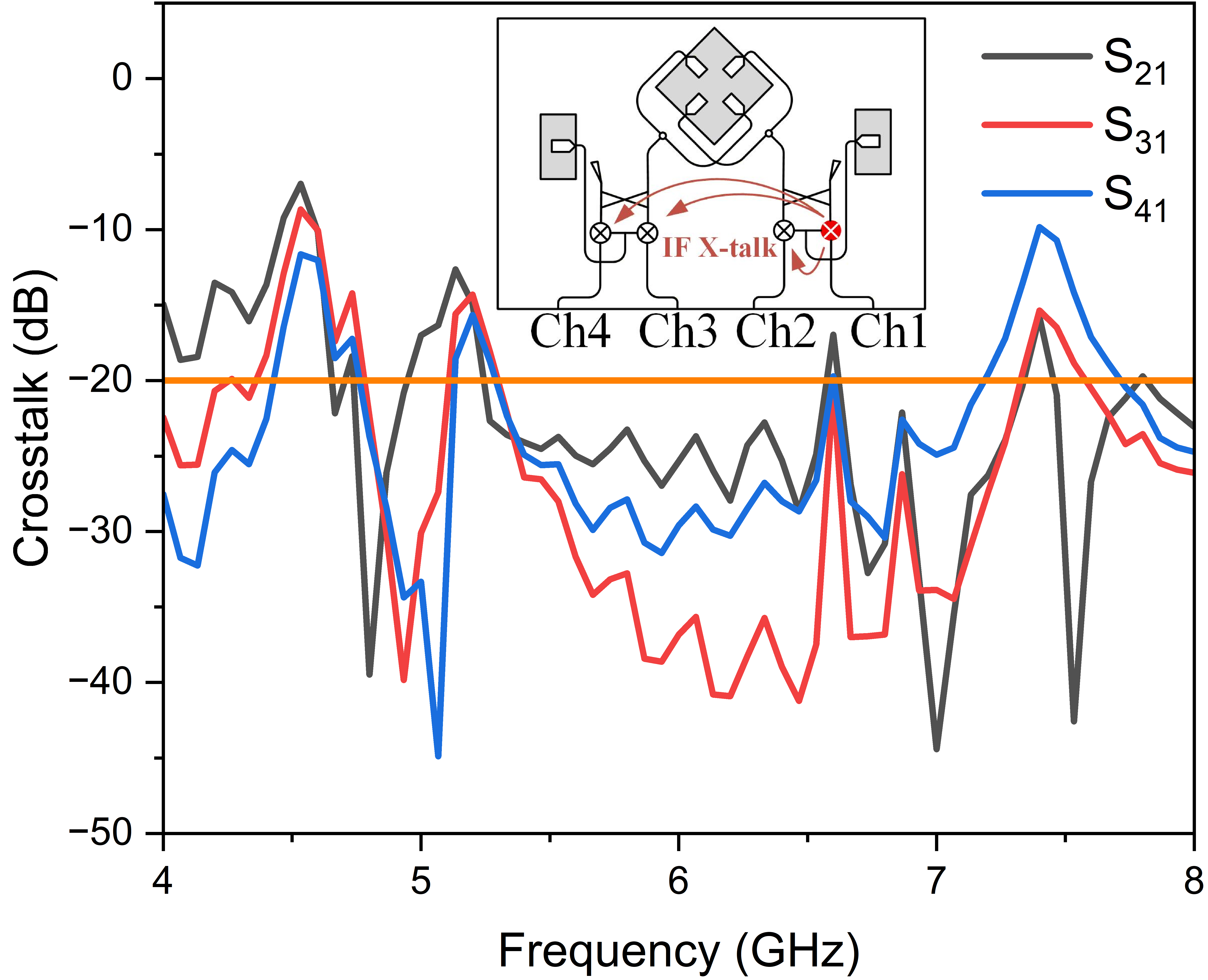}
    \caption{ }
  \end{subfigure}
  % second figure-------------------------
  \begin{subfigure}{0.4\textwidth}
    \centering
    \includegraphics[width=0.8\textwidth,clip]{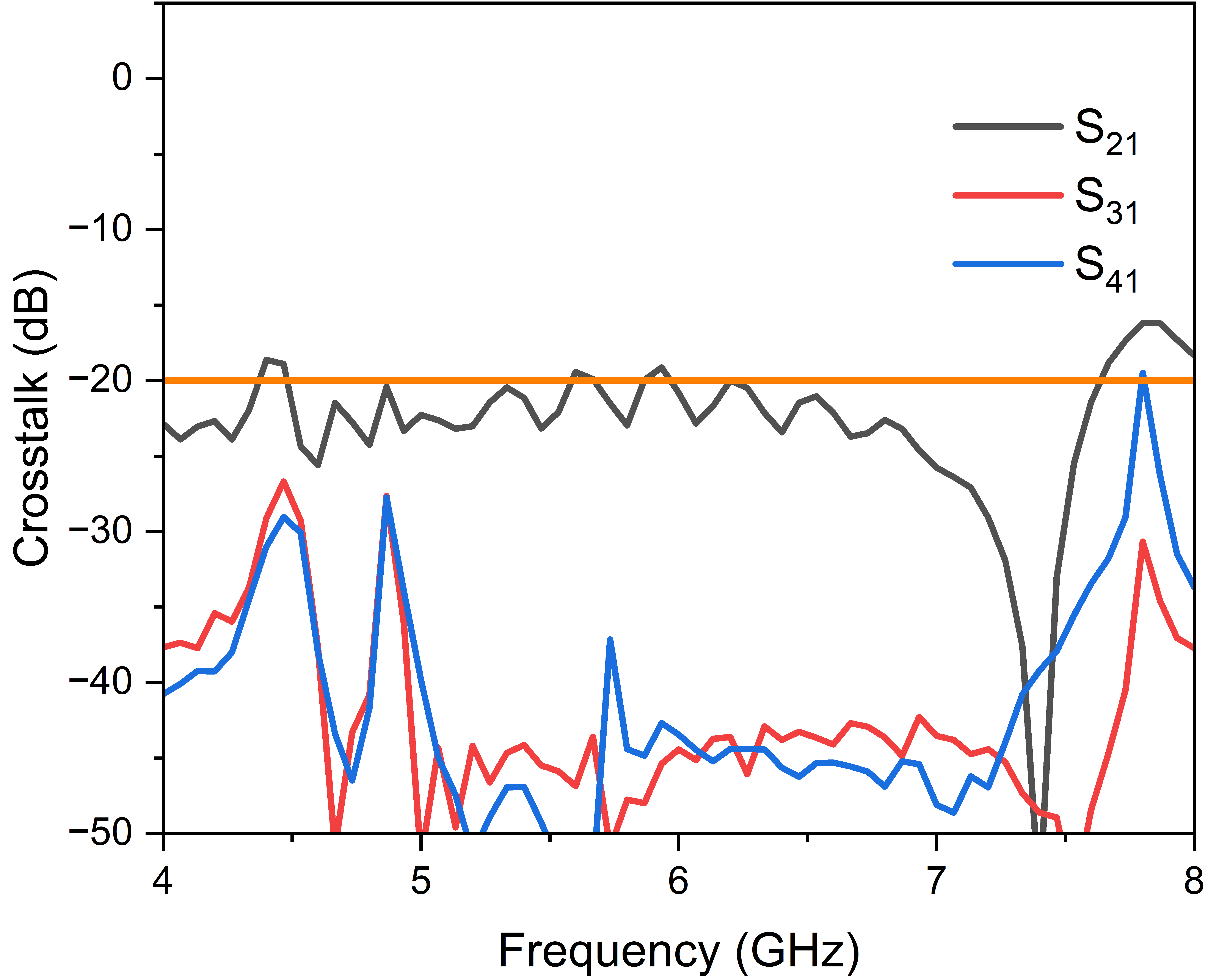}
    \caption{ }
  \end{subfigure}
\caption{Measured inter-channel IF crosstalk before (a) and after (b) the mixer-holder modification shown in Fig.~\ref{FigFieldMatch}. A CW RF signal was injected to generate an IF probe tone only in Ch.~1, while the other channels were biased above the gap voltage so that their mixing efficiency became negligible. Under this condition, the detected signals in Ch.~2, Ch.~3, and Ch.~4 represent IF crosstalk rather than RF mixing. The results are normalized to the signal amplitude of Ch.~1 and corrected for channel-to-channel gain differences. After the modification, the crosstalk level at the resonance frequencies was reduced by more than 10 dB, consistent with the improvement observed in the SBR.}
\label{FigIFXTalk}
\end{figure}

The effectiveness of this solution was first confirmed by achieving SBR greater than 10 dB over most of the RF band, as shown in Fig.~\ref{FigSBR}. This demonstrates that practical sideband separation became possible after suppressing the IF crosstalk. A more direct verification was performed using the measurement scheme illustrated in the inset of Fig.~\ref{FigIFXTalk}(a), where crosstalk between IF channels was measured explicitly. A CW RF signal was injected through the feed horn so that an IF probe tone was generated only in Ch.~1, while the SIS mixers in the other channels were biased far above the gap voltage on the linear branch of the I–V curve, where their mixing efficiency is negligibly small. Under this condition, any detected signal in the other channels originates predominantly from IF crosstalk rather than from RF mixing. The probe frequency was swept across the IF band, and the responses in the other channels were normalized to the amplitude of Ch.~1 and corrected for channel-to-channel gain differences. The measured results before and after the mixer-holder modification are compared in Fig.~\ref{FigIFXTalk}. The crosstalk level was reduced by more than 10 dB at the resonance frequencies, in good agreement with the improvement observed in the SBR measurement.

\section{Conclusion and Outlook}

In this work, we have demonstrated, for the first time, a monolithic dual-polarization 2SB SIS mixer implemented on a silicon-based MMIC platform at 2-mm wavelengths. The presented device successfully integrates the key functionalities required for dual-polarization and sideband-separating operation within a single chip, achieving a sideband rejection ratio exceeding 10 dB across most of the RF band and a minimum SSB receiver noise temperature of approximately 60 K. These results verify the feasibility of extending superconducting MMIC technology toward highly integrated heterodyne receivers.

The study also clarifies several critical technical aspects associated with MMIC-based SIS mixers. A thin-film resistor process has been established to realize lumped-element matched loads required for 2SB circuits. A unique IF crosstalk mechanism caused by substrate cavity modes has been identified as a major limitation for sideband separation. An effective mitigation strategy based on modifying the electromagnetic boundary condition at the MMIC–holder interface has been demonstrated, leading to a significant improvement in sideband rejection.

Despite these achievements, several issues remain to be addressed. The systematic increase in noise temperature at higher IF frequencies suggests an intrinsic limitation associated with the present MMIC architecture that requires further investigation. In addition, the relatively low conversion gain indicates that further optimization of embedding impedance and mixer circuit design is necessary. From a fabrication perspective, the implementation of conductive TSVs is expected to provide a more fundamental solution for suppressing substrate modes and improving channel isolation. Based on the present experimental results, these remaining limitations are considered to be engineering challenges rather than fundamental barriers.

Looking forward, the demonstrated architecture provides a practical path toward large-format heterodyne arrays based on hybrid planar integration at millimeter wavelengths. Furthermore, because the membrane-based on-chip probes are expected to remain applicable at higher frequencies and the measured losses of CPW and microstrip lines do not increase significantly per wavelength toward submillimeter bands \cite{shan2024separation}, this MMIC-based SIS architecture is expected to be extendable to submillimeter wavelengths up to the Nb gap frequency. At submillimeter wavelengths, where atmospheric opacity becomes increasingly severe, the improvement in observational efficiency provided by multibeam heterodyne receivers is expected to become even more significant for ground-based radio telescopes.

%\section*{Acknowledgment}

% Can use something like this to put references on a page
% by themselves when using endfloat and the captionsoff option.
\ifCLASSOPTIONcaptionsoff
  \newpage
\fi

\bibliographystyle{IEEEtran}

\bibliography{MMIC2SBV1}

\end{document}